\documentstyle[aps,prl,twocolumn,epsf,floats]{revtex}
\newcommand{\beq}{\begin{equation}}
\newcommand{\eeq}{\end{equation}}
\newcommand{\bea}{\begin{eqnarray}}
\newcommand{\eea}{\end{eqnarray}}

\newcommand{\veps}{\varepsilon}
\newcommand{\etal}{{\it et al.},}

\begin{document}

\title{\bf \LARGE
Spatial structure of a vortex in low density neutron matter
}
\vspace{0.75cm}
\author{Yongle Yu and Aurel Bulgac}
\vspace{0.50cm}
\address{Department of Physics, University of
Washington, Seattle, WA 98195--1560, USA}

\maketitle

\begin{abstract}

We study in a fully selfconsistent approach the structure of a vortex
in low density superfluid neutron matter. We determine that the matter
density profile of a vortex shows a significant depletion in the
region of the core, a feature never reported for a vortex state in a
Fermi superfluid.

\end{abstract}

\draft
\pacs{PACS numbers: 26.60.+c, 21.65.+f, 97.60.Jd }



Almost thirty years ago Anderson and Itoh \cite{pwa} put forward the
idea that vortices should appear in neutron stars and that they can
also get pinned to the solid crust. They argued that the
"star--quakes," observable on Earth as pulsar "glitches," apparently
the only evidence so far that solid matter exits in the universe apart
from planets, are caused by the vortex de--pinning. This idea and its
various implications have been examined by numerous authors, see
Refs. \cite{aads} and further references therein, but a general
consensus does not seem to have emerged so far. In spite of this,
there does not seem to be a doubt in anybody's mind that a significant
part of the matter in a neutron star is a superfluid of one kind or
another. In particular, for densities less than nuclear saturation
densities one expects neutrons to form $^1S_0$ Cooper pairs, similar
to ordinary electrons in superconductors. Even though at low neutron
densities the kinetic energy dominates over interaction energy, since
this is mainly attractive, Cooper instability occurs. Neutrons become a
Fermi superfluid and vortices can appear. In a low density neutron
matter various nuclear--like objects \cite{nuclei} exist as well, a
state of matter often referred to as "pasta," "meat balls" and "Swiss
cheese" phases. The nuclei--like objects in some of these phases are
significantly denser than the surrounding low density neutron matter
and positively charged and are expected to form a rather stiff Coulomb
lattice. However, other alternatives (liquid crystals, amorphous,
disordered or heterogeneous state) have been discussed in the
literature as well \cite{rigid}. In any case, these inhomogeneities
are the "islands" on which vortices can get "pinned down," at least
temporarily. The estimates of the de--pinning energy available in
literature are rather crude. The profile of a vortex in neutron matter
is typically determined using a Ginzburg--Landau equation, which is
expected to give mostly a qualitative picture and its accuracy is
difficult to estimate.  Surprisingly, there exists only one
microscopic calculation of a vortex in low density neutron matter
\cite{deblasio}.  In regular superconductors the pairing energy is a
relatively small quantity, when compared to the Fermi energy, and the
presence or absence of the pairing field is not expected to lead to
any noticeable changes in the electron density \cite{electrons}.  For
this reason is typical in the case of electrons to neglect the
mean--field (more exactly, to assume it a constant), and only account
for the presence of the magnetic field induced by the electron
super--flow. It is known that inside a vortex core one can have bound
states within the pairing gap \cite{corestates}, basically due to the
phenomenon known as Andreev reflection. The presence of these
localized states inside the vortex core leads to a number of
observables, in particular their presence modifies the specific heat
at low temperatures. This aspect could prove crucial for the thermal
emission from neutron stars, but apparently so far this aspect has not
been considered.  A small change in the profile of the matter density
and of the mean--field was recently reported in dilute atomic Fermi
gases, but it was not considered noteworthy \cite{nicolai}, see also
Ref. \cite{deblasio}.

We report here on a fully microscopic analysis of a vortex in pure
neutron matter and but we do not consider explicitly so far an
impurity, on which such a vortex might get pinned. The ratio of the
pairing gap to the Fermi energy in low density neutron matter
is larger than in other systems and as we have found, the matter
distribution is influenced appreciably by the presence of a vortex. It
is natural to expect that a major change in the density profile of a
vortex is bound to affect the magnitude and the character of the
pinning mechanism. This particular aspect, apparently never reported
in literature until now, can have significant consequences on the
physics of vortices in neutron stars.

\begin{figure}[tbh]
\begin{center}
\epsfxsize=6.0cm
\centerline{\epsffile{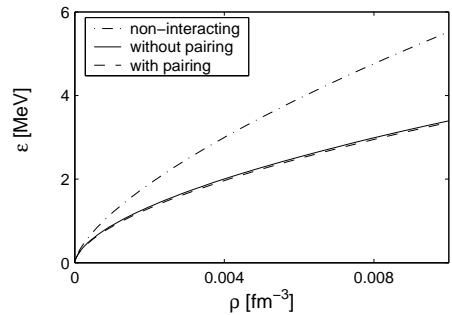}}
\end{center}
\caption{ The energy per particle as a function of neutron density for
the non--interacting neutron gas, interacting with and without pairing
correlations taken into account, for the case of homogeneous matter
distribution. }

\label{fig:fig01}
\end{figure}

\begin{figure}[tbh]
\begin{center}
\epsfxsize=6.0cm
\centerline{\epsffile{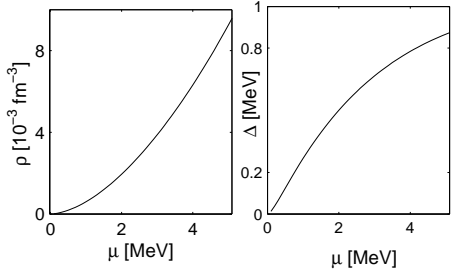}}
\end{center}
\caption{ The neutron density and neutron pairing gap as a function of
the neutron chemical potential in the case of homogeneous neutron
distribution.  }

\label{fig:fig02}
\end{figure}

In order to describe selfconsistently neutron matter, we use a nuclear
energy density functional approach (NEDF), which works surprisingly
well in finite nuclei \cite{abyy,ab,snpb}. This NEDF has as the main
part the contribution describing normal nuclear matter \cite{fayans},
which was fitted by Fayans to reproduce the microscopic many body
calculations of Refs. \cite{nm} for the case of homogeneous matter
distribution. Fayans has supplemented this input with a
phenomenological gradient correction energy density (which we take
into account here) as well as with a spin--orbit energy density
(neglected here).  A separate contribution to the energy density
functional should describe pairing correlations.  It is well known
that even in very dilute systems, the induced interactions play a
significant role in determining the magnitude of the pairing gap,
drastically reducing it by a factor close to 2.2 when compared to a
naive BCS calculation with bare interaction \cite{gor,hh,spr}. Neutron
matter becomes dilute in the sense that $\rho |a|^3 < 1$ (where $a$ is
the nn--scattering length) only at densities smaller than $\approx 6
\times 10^{-6}$ fm$^{-3}$. The range of neutron densities we are
considering here are significantly larger and the arguments presented
in Refs. \cite{gor,hh,spr} strictly do not apply.  Over the years a
number of investigators arrived at qualitatively similar conclusions,
even though the specific corrections considered often varied from one
set of authors to another \cite{clark} and the neutron matter was
certainly not dilute in the sense mentioned above.  One should note
also, that in nuclei \cite{tera} the effect seems to be the opposite,
a conclusion, which for the wrong reasons, agrees with the analysis of
Ref. \cite{hh} for the case when the number of fermion species is four
(two for spin and two for isospin). All analyses however seem to agree
that the pairing gap in neutron matter as a function of the Fermi
momentum has a maximum of approximately 1 MeV for a
Fermi momentum of $k_F\approx 0.8$ fm$^{-1}$. and that the gap
vanishes for Fermi momenta larger than $\approx 1.5$ fm$^{-1}$. This
shape of the pairing gap could be reproduced rather satisfactorily
with the formula \cite{hsu}
\beq
\Delta = \left ( \frac{2}{e} \right ) ^{7/3} \frac{\hbar ^2 k_F^2}{2m}
\exp \left ( -\frac{\pi}{2 \tan \delta (k_F)}\right ),
\eeq
where $k_F$ is the Fermi wave vector, $m$ the nucleon mass and
$\delta(k_F)$ is the $^1S_0$ nn--phase shift. Following the arguments
given in Refs. \cite{gor,hh}, we have changed the corresponding
formula for the gap given in Ref. \cite{hsu} by including the effect
of induced interactions as an additional factor of
$1/(4e)^{1/3}\approx 0.45$.  Since for the neutron densities we are
considering here, the effective range approximation to the Nijmegen
$^1S_0$ nn--phase shift is essentially exact \cite{nn}, we have
parametrized the pairing gap in homogeneous neutron matter using this
approximation. The resulting pairing gap is well within the
theoretical uncertainties, including various corrections as discussed in
Refs. \cite{clark} and references therein. For each value of the
neutron density we have determined a corresponding value for the
``bare coupling constant'' $g(\rho)$ as discussed in
Refs. \cite{ab}. This completes the construction of the NEDF, in
complete analogy with the construction of the normal part of the
electron EDF \cite{hk}, see
Figs. \ref{fig:fig01}--\ref{fig:fig02} for details.

\begin{figure}[tbh]

\begin{center}

\epsfxsize=6.0cm

\centerline{\epsffile{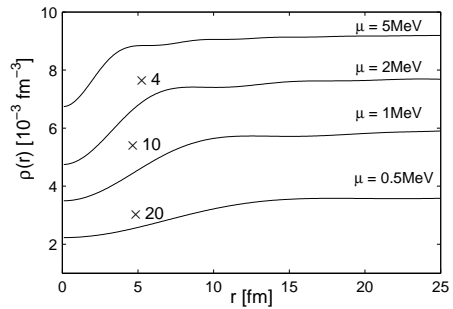}}

\end{center}

\caption{ The neutron density as a function of the distance from the
vortex axis $r$.  Some of these density distributions have been
rescaled by the factor indicated above the corresponding curve. }

\label{fig:fig1}
\end{figure}

\begin{figure}[tbh]
\begin{center}
\epsfxsize=6.0cm
\centerline{\epsffile{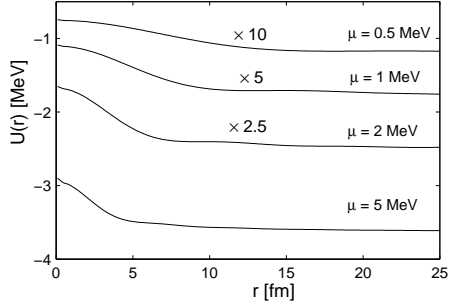}}
\end{center}
\caption{ The mean--field neutron potential as a function of the
distance from the vortex axis. Some of these potentials have been
rescaled by the factor indicated above the corresponding curve.  }

\label{fig:fig2}
\end{figure}

The rest of the technical details were described previously in
Refs. \cite{abyy,ab,snpb}.  We note only that the normal and anomalous
densities were evaluated using the technique initially described in
Ref. \cite{stb}, and consequently all the boundary conditions were
taken into account exactly and there was no need to either enclose the
vortex in a cylinder or to introduce periodic boundary conditions
along the vortex axis. Moreover, there is no need to diagonalize any
matrices and the fact that the single--quasiparticle spectrum has (or
has not) a number of discrete eigenvalues does not require any change
in the numerical algorithm.  Since we consider a straight vortex, the
quasiparticle wave functions in cylindrical coordinates
$\bbox{r}=(r,\phi,z)$ have the simple structure

\beq
\left (
\begin{array}{l}
u_{\bbox{\alpha}}(\bbox{r})\\ v_{\bbox{\alpha}} (\bbox{r})
\end{array}
\right )
=\left (
\begin{array}{l}
u_\alpha(r)\exp\left [ in\phi +  ikz\right ] \\
v_\alpha(r)\exp\left [ i(n-1)\phi + ikz\right ]
\end{array}
\right ),
\eeq
where $\alpha$ labels the quasiparticle states, $n$ is integer and
$k$ is the wave vector of the quasiparticle state along the vortex
$Oz$--axis and $\bbox{\alpha}=(\alpha,k,n)$.
(Note that we use $\bbox{r}$ for the 3d--coordinate and
$r$ for the radial cylindrical coordinate.) The mean--field depends only
on radial coordinate $r$ (measured from the vortex axis), while the
pairing field has the structure $\Delta (r)\exp(i\phi)$.

\begin{figure}[tbh]
\begin{center}
\epsfxsize=6.0cm
\centerline{\epsffile{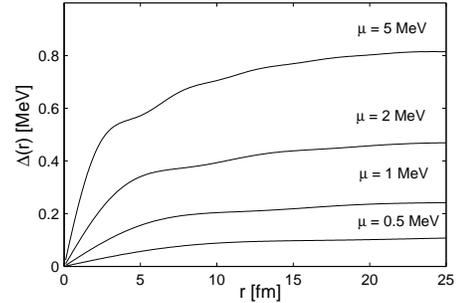}}
\end{center}
\caption{ The pairing gap as a function of the distance from the
vortex axis.  }

\label{fig:fig3}
\end{figure}

\begin{figure}[tbh]
\begin{center}
\epsfxsize=6.0cm
\centerline{\epsffile{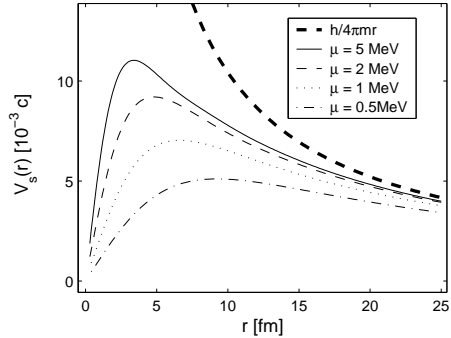}}
\end{center}
\caption{ The neutron current velocity $V_s(r)$ (in units of speed of
light) as a function of the distance from the vortex axis. The thick
dashed line is a plot of the ``classical'' current velocity
$V_v(r)=\hbar /2mr$. }

\label{fig:fig4}
\end{figure}

Our results are summarized in Figs. \ref{fig:fig1} -- \ref{fig:fig4}
where we show the vortex density, mean--field, pairing field and
velocity profiles, as functions of the distance from the vortex axis,
for several values of the neutron chemical potential. The most
unexpected feature of these profiles is the prominent depression of
the matter density in the region of the vortex axis. In the case of
superfluid dilute bosons at the vortex axis the density practically
vanishes. Except for the barely visible features in the density
profile of a vortex reported in Refs. \cite{deblasio,nicolai} but not
commented on, such features have not been discussed in literature in
the case of a Fermi system. The asymptotic density changes roughly by
a factor of forty from the lowest to the highest values of the
chemical potential plotted in these figures and the Fermi momentum
changes by approximately a factor of three. The size of the
inhomogeneity in the matter distribution and the mean--field $U(r)$
are governed by the asymptotic value of the Fermi wavelength, while
the spatial profile of the pairing field $\Delta (r)$ is controlled
mostly by the coherence length, $\xi= \varepsilon_F/\Delta k_F\gg
1/k_F$ (where $k_F$, $\Delta$ and $\varepsilon_F$ are the asymptotic
values for the Fermi wave vector, pairing gap and Fermi energy). The
density gradients are noticeable and the gradient correction terms in
the NEDF, even though they are small, are non--negligible. The density
depletion drives the mean--field to become less attractive in the
vortex core, which in its turn, due to the self-consistency,
stabilizes this structure. The density depletion is likely so
significant in neutron matter because the pairing field is relatively
stronger than in the case of electrons. The magnitude of the pairing
correlations is characterized by the ratio $\Delta(r)/\veps_F
(r)=\Delta (r)/(\mu-U(r))$. In the vortex core region the behavior
$\Delta(r) \rightarrow 0$ as $r\rightarrow 0$ can be partially
compensated by an increasing $U(r)$, which induces in its turn a
density depletion. The density and pairing field profiles (and to a
lesser extent the mean--field as well) show some rather faint
Friedel--like oscillations, which can be attributed to the presence of
discrete states inside the vortex core
\cite{deblasio,electrons,corestates,nicolai}.

The presence of the vortex implies a net flow of the neutron
superfluid around the vortex axis with the velocity given by
\beq
\mathrm{V}_s(r)\bbox{\hat{e}_\phi}
=-\frac{i\hbar}{mr \rho(r) }\sum_{\bbox{\alpha}}
v^*_{\bbox{\alpha}} (\bbox{r})
\bbox{\hat{e}_\phi}\frac{\partial}{\partial \phi}
v_{\bbox{\alpha}}(\bbox{r}),
\eeq
where $\bbox{\hat{e}_\phi} =(-y,x,0)/r$, the summation over
$\bbox{\alpha}$ should be interpreted as a sum or integral when
appropriate, over all quasiparticle states with energies $E_\alpha>0$
and $\rho(r)$ is the neutron number density. Naturally
$\mathrm{V}_s(0)=0$.  If all neutron matter would be involved in such
a super--flow around the vortex axis, the superfluid velocity would be
given by
\beq
\mathrm{V}_v(r)\bbox{\hat{e}_\phi} =\frac{\hbar}{2mr^2}(-y,x,0),
\eeq
which corresponds to Osanger's quantization condition \cite{osanger}
per Cooper pair
\beq
\frac{1}{2\pi} \oint _C \mathrm{V}_v(r)\bbox{\hat{e}_\phi}\cdot d\bbox{r}
  =\frac{\hbar}{2m},
\eeq
and where the contour $C$ is any closed curve, around the vortex axis.
While at relatively large distances from the vortex axis the neutron
super--flow indeed approaches this ``classical'' limit, that is not
the case at distances smaller or of the order of the coherence length
$\xi$. The interpretation of this behavior is relatively simple. Since
the pairing field vanishes at the vortex axis, the neutron matter in
the immediate neighborhood is not fully superfluid. Only the fraction
$f_s(r)=\mathrm{V}_s(r)/\mathrm{V}_v(r)$ of the neutrons are entrained
into the superfluid flow. This fraction vanishes at the vortex axis
and approaches unity far from the vortex core, essentially in a
monotonic fashion. $\mathrm{V}_s(r)$ is always significantly smaller
than the Fermi velocity.

The existence of a strong density depletion in the vortex core is
going to affect appreciably the energetics of a neutron star crust.
One can obtain a gross estimate of the pinning energy of
a vortex on a nucleus as $E^V_{pin}=[\veps(\rho_{out}) \rho_{out} -
\veps(\rho_{in}) \rho_{in}]V$, where $\veps(\rho)$ is the energy per
particle at density $\rho$, see Fig. \ref{fig:fig01}, $\rho_{in}$ and
$\rho_{out}$ are the densities inside and outside the vortex core and
$V$ is the volume of the nucleus. Naturally, this simple formula does
not take into account a number of factors, in particular surface
effects and the changes in the velocity profile and the pairing
field. These last contributions were accounted for (with some
variations) in the past \cite{pwa,aads}. However, if the density
inside the vortex core and outside differ significantly one expects
$E^V_{pin}$ to be the dominant contribution.  In the low density
region, where $\veps(\rho_{out})\rho_{out}/\veps(\rho_{in})\rho_{in}$
is largest, see Fig. \ref{fig:fig2}, one expects a particularly large
anti--pinning effect $(E^V_{pin}>0)$. The energy per unit length of a
simple vortex is expected to be significantly lowered when compared
with previous estimates \cite{pwa,aads} by $\approx [\veps(\rho_{out})
\rho_{out} - \veps(\rho_{in}) \rho_{in}]\pi R^2$, where $R$ is an
approximate core radius, see Fig. \ref{fig:fig1}.
How a significant density depletion found here affects the neutron
star properties in the crust will be addressed in our future work.

This work benefited from partial financial support from DOE under
contract DE--FG03--97ER41014.


\end{document}